# White matter biomarkers from fast protocols using axially symmetric diffusion kurtosis imaging

Brian Hansen[1]*, Ahmad R. Khan[1], Noam Shemesh[2], Torben E. Lund[1], Ryan Sangill[1], Simon F. Eskildsen[1], Leif Østergaard[1], Sune N. Jespersen[1,3]

Author affiliations

[1] Center of Functionally Integrative Neuroscience (CFIN) and MINDLab, Department of Clinical Medicine, Aarhus University, Aarhus, Denmark.

[2] Champalimaud Neuroscience Programme, Champalimaud Centre for the Unknown, Lisbon, Portugal

[3] Department of Physics and Astronomy, Aarhus University, Aarhus, Denmark.

* Corresponding Author

Brian Hansen, CFIN, Aarhus University

Building 10G, 5th Floor, Nørrebrogade 44, DK-8000 Århus C, Denmark

Email: brian@cfin.au.dk

**Abbreviations used:**

AWF: axonal water fraction

aWMTI: axial white matter tract integrity

CSF: cerebrospinal fluid

DDE: Double diffusion encoding

DKI: Diffusion kurtosis imaging

DTI: Diffusion tensor imaging

DW: diffusion weighted

EM: electron microscopy

EPI: echo planar imaging

EU: European Union

faWMTI: fast axial white matter tract integrity

KFA: kurtosis fractional anisotropy

MK: mean kurtosis

MKT: mean of the kurtosis tensor

PBS: phosphate buffered saline

PFA: paraformaldehyde

PGSE: pulsed gradient spin echo

RK: radial kurtosis

SC: spinal cord

SDE: single diffusion encoding

SNR: signal to noise ratio

WM: white matter

WMTI: white matter tract integrity

# Abstract

White matter tract integrity (WMTI) can characterize brain microstructure in areas with highly aligned fiber bundles. Several WMTI biomarkers have now been validated against microscopy and provided promising results in studies of brain development and aging, as well as in a number of brain disorders. Currently, WMTI is mostly used in dedicated animal studies and clinical studies of slowly progressing diseases but has not yet emerged as a routine clinical tool. To this end, a less data intensive experimental method would be beneficial by enabling high resolution validation studies, and ease clinical applications by speeding up data acquisition compared to typical diffusion kurtosis imaging (DKI) protocols utilized as part of WMTI imaging.

Here, we evaluate WMTI based on recently introduced axially symmetric DKI which has lower data demand than conventional DKI. We compare WMTI parameters derived from conventional DKI to those calculated analytically from axially symmetric DKI. We employ numerical simulations, as well as data from fixed rat spinal cord (1 sample) and in vivo human (3 subjects) and rat brain (4 animals). Our analysis shows that analytical WMTI based on axially symmetric DKI with sparse data sets (19 images) produces WMTI metrics that correlate strongly with estimates based on traditional DKI data sets (60 images or more). We demonstrate the preclinical potential of the proposed WMTI technique in in vivo rat brain (300 µm isotropic resolution with whole brain coverage in a one hour acquisition). WMTI parameter estimates are subject to a duality leading to two solution branches dependent on a sign choice which is currently debated. Results from both of these branches are presented and discussed throughout our analysis. The proposed fast WMTI approach may be useful for preclinical research and e.g. clinical evaluation of patients with traumatic white matter injuries or symptoms of neurovascular or neuroinflammatory disorders.

# Introduction

Diffusion kurtosis imaging (DKI) (1,2) quantifies the leading deviation from Gaussian diffusion. Since this deviation is a consequence of tissue microstructure's influence on the water diffusion profile, DKI has increased sensitivity to tissue microstructure over diffusion tensor imaging (DTI). DKI yields a number of parameters that have been shown to be sensitive to subtle changes in brain tissue organization, either as a consequence of disease such as demonstrated for stroke (3,4), Alzheimer's disease (5), multiple sclerosis (6), head trauma (7-9) (and reviewed in (10)), or natural effects such as development and aging (11,12). This sensitivity is present in both gray and white matter (WM). However, in WM, more information is available when diffusion and kurtosis tensors are used in combination with modeling as in the white matter tract integrity (WMTI) technique (13). In suitable WM regions (typically selected based on indices describing the diffusion tensor shape), WMTI extracts detailed information about WM fiber bundle composition, such as intra- and extra-axonal diffusivities, axonal water fraction (AWF), and the tortuosity, $\alpha$, of the extra-axonal space.

The WMTI technique has been applied to normal brain development and aging (14,15) as well as diseases including Alzheimer's disease (16,17), mild head trauma (18), multiple sclerosis (19), autism (20), and stroke (3). More recently, DKI based indices of WM microstructure have compared favorably with histology and electron microscopy (21-24) confirming the ability of WMTI to extract microstructural changes in highly aligned WM. A recent study found WMTI to have broader applicability than other DKI-based tissue models (25). However, the widespread clinical application of WMTI awaits not only a firm understanding of its cytoarchitectural correlates, but also ways of reducing acquisition and image processing time demands, in that WMTI requires a full DKI acquisition, typically including 60 images (13) or more (26), and computationally demanding post-processing.

Recently, a strategy for reducing the number of parameters in the DKI signal expression from 22 to 8 was proposed and evaluated in the context of fast estimation of radial and axial kurtosis (27). This simplification of the signal expression was achieved by imposing axial symmetry on both the diffusion and kurtosis tensors. This symmetry assumption was shown to have little effect on directional DKI parameter estimates, even in regions unlikely to have axial symmetry in reality and when the axially symmetric tensors were determined from small data sets containing only 19 images.

The primary aim of this study is to evaluate axisymmetric DKI as a strategy for reducing the data requirement for WMTI. With fewer parameters to determine, axisymmetric DKI might allow acceleration of the data acquisition for WMTI which would alleviate some of the experimental challenges currently associated with validation, clinical exploration and application. For instance, in preclinical studies, accelerated WMTI would allow acquisition of data sets with higher spatial resolution or higher signal to noise ratio (SNR) (more averages) in the same amount of time as a traditional DKI acquisition for WMTI. In the clinic, the major benefit of such rapid acquisitions would be more widespread use and exploration of WMTI. The smaller data requirement might also be a benefit for imaging in critically ill patients, or where the subject has difficulty lying still during a lengthy examination, e.g. the young and elderly. Short WMTI acquisitions may also be useful in body DKI where motion is often a more severe problem, for example for evaluation of spinal cord (SC) or peripheral nerve.

The most recent version of WMTI does not assume strict axial symmetry (13). Then, the axonal water fraction is determined by numerically maximizing the apparent kurtosis over a large number of directions in each pixel followed by maximization over a local neighborhood of (assumed) homogenous tissue. By assuming axial symmetry, we can employ a faster strategy, by which all parameters can be analytically determined from mean, axial and radial diffusivity and kurtosis. This WMTI estimation strategy is similar to an earlier version of WMTI (28), which however, did not employ axially symmetric DKI as is done here.

We evaluate the parameter fidelity of these strategies through simulations and compare our WMTI estimates to a (simplified) implementation of conventional WMTI using experimental data from a range of systems. We also investigate estimation precision when based on either large or small data sets made possible by the compact axisymmetric DKI signal representation. Rapid, optimal protocols for accelerated WMTI based on axisymmetric DKI may exist but are not the subject of this study. Instead we focus on data acquired with the 1-9-9 strategy for fast kurtosis imaging (27,29-31) and assess performance of the analytical WMTI variant based on such acquisitions. In the first part of our analysis, numerical simulations are used to compare the performance of conventional WMTI to the proposed analytical strategy based on axially symmetric DKI when axial symmetry is not fulfilled. We then analyze high resolution data from rat SC acquired at 16.4T, and data from three human volunteers acquired at 3T. These data acquisitions are constructed to allow analysis with WMTI as well as the analytical variants (either based on a large data set or a small subset of 19 images), for direct method comparison. Finally, we present examples of analytical WMTI based on 19 images acquired at 9.4T in four rats *in vivo* with whole brain coverage at an isotropic resolution of 300 µm. This serves to illustrate the preclinical potential of the proposed methods. Based on our analysis we discuss parameter behavior in the two solution branches produced by an inherent sign ambiguity related to the relative magnitude of intra-and extra-axonal diffusivity. This duality is currently a matter of interest (13,32) and we therefore provide histograms of the affected diffusivities from both branches in all investigated systems and discuss branch estimates in the context of their expected physical parameter behavior.

# Theory

*DKI*

The standard expression for the DKI signal is (1):

$$\log S(b,\hat{n})/S_0 = -bn_i n_j D_{ij} + \frac{1}{6}b^2 \bar{D}^2 n_i n_j n_k n_l W_{ijkl} = -bD(\hat{n}) + \frac{1}{6}b^2 \bar{D}^2 W(\hat{n})$$
$$= -bD(\hat{n}) + \frac{1}{6}b^2 D(\hat{n})^2 K(\hat{n})$$
(1)

Here $S_0 = S(b=0)$ is the signal normalization and $b$ is the diffusion weighting applied along a direction $\hat{n} = (n_x, n_y, n_z)$. In this notation, subscripts label Cartesian components (e.g. $i = x,y,z$) and Einstein notation for summation is used. D denotes the diffusion tensor (33), and definitions of the kurtosis tensor W and apparent excess kurtosis $K(\hat{n})$ are adopted from (1). Here, as in the rest of the paper, overbar denotes mean value i.e. $\bar{D}$ is mean diffusivity. Typical DKI acquisitions contain about 60 images in total: 30 directions at two non-zero b-values (typically 1.0 ms/µm² and 2.0-2.5 ms/µm²) and a few additional unweighted images for signal normalization, which are used to estimate D and W by fitting to Eq. (1). Typically, DKI is acquired with single diffusion encoding (SDE) (34), but double diffusion encoding (DDE) DKI has been considered (35,36).

*WMTI*

From D and W, metrics of white matter tract integrity (WMTI) can be derived based on modeling described in (13). The WMTI model describes the intra-axonal space as sticks with an effective radius of zero (valid for clinical diffusion gradients, see discussion). Furthermore, water is assumed not to exchange between the intra- and extra-axonal spaces. Diffusion in each compartment is approximated with anisotropic Gaussian diffusion. Although individual axons behave as sticks for clinical diffusion times, the presence of multiple non parallel populations (or dispersion) may cause the intra-axonal tensor to have three non-vanishing eigenvalues (13). Hence, the signal expression for this two-compartment system becomes:

$$S(b,\hat{n})/S_0 = f \exp(-b\hat{n}^T D_a \hat{n}) + (1-f)\exp(-b\hat{n}^T D_e \hat{n})$$
(2)

where $D_a$ and $D_e$ are diffusion tensors belonging to the intra-axonal and extra-axonal space respectively, $f$ is the MR visible axonal water fraction (AWF), and the intra- and extra axonal diffusion tensors are (subscript numbers refer to sorting of eigenvalues by magnitude):

$$\mathbf{D}_a = \begin{bmatrix} D_{a3} & 0 & 0 \\ 0 & D_{a2} & 0 \\ 0 & 0 & D_{a1} \end{bmatrix}, \qquad \mathbf{D}_e = \begin{bmatrix} D_{e3} & 0 & 0 \\ 0 & D_{e2} & 0 \\ 0 & 0 & D_{e1} \end{bmatrix} \qquad (3)$$

From the cumulant expression of this model, the measured diffusion and kurtosis tensors can be expressed in terms of the model parameters. Based on these relations, WMTI provides estimates of the (MR visible) axonal water fraction (AWF), axonal diffusivity ($D_a = \mathrm{Tr}(\mathbf{D}_a)$), extra-axonal parallel and radial diffusivities ($D_{e,\parallel} = D_{e,1}$ and $D_{e,\perp} = (D_{e2} + D_{e3})/2$), and the tortuosity ($\alpha = D_{e,\parallel}/D_{e,\perp}$) of the extra-axonal space, all obtained from a DKI data set where AWF estimation would otherwise require significantly higher b-values (13,37). If the axonal space consists of more than one fiber population the framework was shown to be robust for small intersection angles (<30°) because such axonal arrangements look essentially Gaussian for typical a DKI protocol (13). The estimation procedure relies on optimizing a lower bound, i.e. an inequality rather than an equality, which is carried out by numerically maximizing the apparent kurtosis $K(\hat{n})$ over a large number of directions. If the fiber bundles are non-coplanar, an improved estimation method additionally entails averaging over a local neighborhood of homogeneous tissue.

*Axially symmetric DKI*

All previous WMTI studies have used the full representation of the DKI signal in Eq. (1) for estimation of D and W, without any symmetry forced on the tensors (22 parameters) (1,2). Recently it was shown (27), that the apparent kurtosis $W(\hat{n})$ can be expressed by only three independent parameters if one assumes the system to possess axial symmetry. This simplification was found to yield reliable estimates of directional kurtosis and diffusion metrics even if axial symmetry is unlikely to apply in reality. Choosing $\hat{z}$ to be parallel to the axis of symmetry, $W(\hat{n})$ is characterized by $\overline{W}$ (29), $W_{\parallel}$ (parallel kurtosis, the apparent kurtosis along $\hat{z}$) and $W_{\perp}$ (radial kurtosis, apparent kurtosis along any direction in the *xy*-plane). The axis of symmetry must be specified as well (two angles with respect to the lab frame), resulting in a total of only 5 degrees of freedom for an axially symmetric kurtosis tensor. The axially symmetric diffusion tensor shares the symmetry axis of W and hence adds only two parameters. With normalization, the number of parameters for axially symmetric DKI therefore adds up to eight (27).

In this axially symmetric system, the apparent kurtosis along an arbitrary direction is characterized by the polar angle $\theta$:

$$W(\theta) = \frac{1}{16}\left(\cos(4\theta)(10W_\perp + 5W_\| - 15\overline{W}) + 8\cos(2\theta)(W_\| - W_\perp) - 2W_\perp + 3W_\| + 15\overline{W}\right) \quad (4)$$

and similarly for the apparent diffusivity, $D(\theta)$:

$$D(\theta) = D_\perp + \cos^2(\theta)(D_\| - D_\perp) \quad (5)$$

If the direction of diffusion weighting is not well defined, e.g. when imaging gradients add significant cross terms, explicit, coordinate independent tensor forms are required:

$$\mathrm{W} = \frac{1}{2}(10W_\perp + 5W_\| - 15\overline{W})\mathrm{P} + W_\perp \mathbb{I} + \frac{3}{2}(5\overline{W} - W_\| - 4W_\perp)\mathrm{Q} \quad (6)$$

$$\mathrm{D} = D_\perp \mathbb{I} + (D_\| - D_\perp)\mathbf{u}\mathbf{u}^T \quad (7)$$

Here **u** is a unit vector along the axis of symmetry. The definitions of the tensors P, $\mathbb{I}$ and Q are (subscripts again label Cartesian components):

$$\begin{aligned} P_{ijkl} &= u_i u_j u_k u_l \\ Q_{ijkl} &= \frac{1}{6}(u_i u_j \delta_{kl} + u_i u_k \delta_{jl} + u_i u_l \delta_{jk} + u_j u_k \delta_{il} + u_j u_l \delta_{ik} + u_k u_l \delta_{ij}) \\ \mathbb{I}_{ijkl} &= \frac{1}{3}(\delta_{ij}\delta_{kl} + \delta_{ik}\delta_{jl} + \delta_{il}\delta_{jk}) \end{aligned} \quad (8)$$

Both terms needed for computing the signal (see Eq. (1)) can then be calculated:

$$\begin{aligned} b_{ij}b_{kl}W_{ijkl} = &\frac{1}{2}(10W_\perp + 5W_\| - 15\overline{W})(\mathbf{u}^T\mathbf{b}\mathbf{u})^2 + \\ &\frac{1}{2}(5\overline{W} - W_\| - 4W_\perp)(\mathbf{u}^T\mathbf{b}\mathbf{u}\,\mathrm{Tr}(\mathbf{b}) + 2\mathbf{u}^T\mathbf{b}\mathbf{b}\mathbf{u}) + \frac{W_\perp}{3}(\mathrm{Tr}(\mathbf{b})^2 + 2\mathrm{Tr}(\mathbf{b}\otimes\mathbf{b})) \end{aligned} \quad (9)$$

$$b_{ij}D_{ij} = \mathrm{Tr}(\mathbf{b})D_\perp + (D_\| - D_\perp)\mathbf{u}^T\mathbf{b}\mathbf{u} \quad (10)$$

In the above, b is the experimental b-matrix, $\otimes$ denotes the tensor direct product such that for two second order tensors A and B, $\mathrm{A}\otimes\mathrm{B}$ is fourth order with Cartesian components $(\mathrm{A}\otimes\mathrm{B})_{ijkl} = A_{ij}B_{kl}$. Using the expressions in Eqs. (6)-(10), we here estimate the tensors D and W with nonlinear least squares fitting to the signal. In the following we show how the axially symmetric tensors provide parameters that can be used directly to determine the axially symmetric WMTI parameters analytically.

*Analytical expressions for the WMTI parameters for parallel fibers:*

We exploit the relationship found in (2) for 2-compartment Gaussian systems:

$$\begin{aligned} D(\hat{n}) &= fD_a(\hat{n}) + (1-f)D_e(\hat{n}) \\ W(\hat{n})\bar{D}^2 &= 3f(1-f)(D_a(\hat{n}) - D_e(\hat{n}))^2 \end{aligned} \quad (11)$$

Evaluated along the radial and axial directions, and averaged over all directions these yield directional diffusivities and kurtoses from which closed form solutions for $AWF$, $D_a$, $D_{e,\perp}$, and $D_{e,\parallel}$ can be derived:

$$\begin{aligned} D_\perp &= (1-f)D_{e,\perp} & (a) \\ D_\parallel &= fD_a + (1-f)D_{e,\parallel} & (b) \\ W_\perp \bar{D}^2 &= 3f(1-f)D_{e,\perp}^2 & (c) \\ W_\parallel \bar{D}^2 &= 3f(1-f)(D_a - D_{e,\parallel})^2 & (d) \\ \bar{W}\bar{D}^2 &= 3f(1-f)\left[D_{e,\perp}^2 + \frac{1}{15}(D_{e,\parallel} - D_a - D_{e,\perp})(7D_{e,\perp} + 3(D_{e,\parallel} - D_a))\right] & (e) \end{aligned} \quad (12)$$

Here, the mean (of the) kurtosis tensor (MKT or $\bar{W}$) is defined as in (29):

$$\bar{W} = \frac{1}{4\pi}\int_{S_2} d\hat{n}\, W(\hat{n}) = \frac{1}{5}\text{Tr}(\mathbf{W}) \quad (13)$$

Tr is the trace. Radial and axial kurtosis are defined as in (27):

$$\begin{aligned} W_\perp &\equiv \frac{1}{2\pi}\int_{S_1} d\hat{n}\, W(\hat{n}) = \frac{1}{4}\left(W(\mathbf{v}_2) + W(\mathbf{v}_3) + W((\mathbf{v}_2+\mathbf{v}_3)/\sqrt{2}) + W((\mathbf{v}_2-\mathbf{v}_3)/\sqrt{2})\right) \\ W_\parallel &= W(\mathbf{v}_1) \end{aligned} \quad (14)$$

where **v**₁, **v**₂ and **v**₃ are diffusion tensor eigenvectors in decreasing order of the eigenvalues. Hence, the left-hand sides of Eq. (12) are readily determined from both the general DKI fit and axially symmetric DKI.

Equation (12) consists of five equations with only four unknowns, due to the assumption of axons as parallel sticks, leaving a choice for which equation to omit - in fact, only 4 of the equations are independent. Noting that $\bar{W}$ is more robustly estimated from 1-9-9 than $W_\parallel$ (as shown in (27)), we here employ Eq. (12e) rather than Eq. (12d) involving the noisier $W_\parallel$. Equations (12) can then be inverted to find expressions for the WMTI parameters:

$$f = AWF = \left(1 + 3D_\perp^2 / W_\perp \bar{D}^2\right)^{-1} \qquad (a)$$

$$D_{e,\perp} = D_\perp / (1-f) \qquad (b)$$

$$D_{e,\parallel} = D_\parallel - \frac{2}{3}\frac{f}{1-f}\left(D_\perp \pm \sqrt{\frac{15(1-f)}{4f}\bar{D}^2\bar{W} - 5D_\perp^2}\right) \qquad (c) \qquad (15)$$

$$D_a = D_\parallel - \frac{2}{3}\left(D_\perp \mp \sqrt{\frac{15(1-f)}{4f}\bar{D}^2\bar{W} - 5D_\perp^2}\right) \qquad (d)$$

$$\alpha = D_{e,\parallel}/D_{e,\perp} \qquad (e)$$

The notation in Eqs. (15) reflects our implementation and shows the interdependence of the parameters. As in standard WMTI, a sign ambiguity exists, which is rooted in the appearance of diffusion coefficients squared in Eq. (12). This demands a choice between two branches yielding solutions with $D_a \leq D_{e,\parallel}$ or $D_a > D_{e,\parallel}$. Since this cannot be decided without independent information, and is an important topic in the current debate (32,38) we report affected parameters ($D_{e,\parallel}$, $D_a$, and $\alpha$) for both branches. We refer to these branches as Branch 1 (yielding $D_a \leq D_{e,\parallel}$) and Branch 2. Note also that the solutions for branch 2 in axially symmetric WMTI are much more obvious than conventional WMTI, where the choice of sign can depend on diffusion direction.

*Comparison strategies:*

We evaluate two axisymmetric WMTI strategies although more strategies are possible. We compare these methods to conventional WMTI (merely WMTI from here on) based on estimates of D and W obtained from a fit of Eq. (1) to suitable data sets. Our first method is analytically evaluated WMTI based on an axisymmetric DKI fit to data sets identical to those we use for WMTI. We refer to this WMTI strategy as axisymmetric WMTI (aWMTI). Our second method is identical to the first, except it exploits the lower data demand for axisymmtric DKI. With only eight parameters in the axisymmetric DKI signal representation the aWMTI strategy may also be applied to sparse data sets. Optimal protocols may exist, but here we employ the 1-9-9 protocol as an example of a compact kurtosis measurement scheme and for consistency with our previous work (27,31,39). In this protocol, one b=0 image is acquired followed by nine distinct encoding directions at each of two b-values $b_1$ and $b_2$; due to this sampling design we refer to the protocol as the 1-9-9 protocol for fast kurtosis imaging (31). The nine directions needed for this scheme are provided in tables elsewhere (27,31). We refer to this WMTI variant as fast axisymmetric WMTI (faWMTI) where the axisymmetric DKI fit parameters from a 1-9-9 data set are used directly with the analytical expressions. We emphasize, that the axisymmetric WMTI method can be used with any compact DKI acquisition with eight or more data points acquired along non-collinear encoding directions. Our method of evaluation therefore uses a two-step approach comparing first WMTI to aWMTI (to assess the isolated effect on WMTI of imposing axial symmetry on the DKI signal expression) and then aWMTI to faWMTI to show the effect of a reduced data foundation. This allows users interested in employing aWMTI (which has reduced processing time due to the WMTI parameters being evaluated analytically) to conventional DKI data to assess method performance while also showing method performance using a sparse sampling scheme such as 1-9-9 protocol.

## Methods

Throughout, the signal to noise ratio (SNR) at b=0 prior to preprocessing is reported. The SNR was calculated as the average signal in a uniform region in the object divided by the standard deviation (std) in a signal free background region corrected for Rayleigh distribution in a standard fashion.

*16.4T MRI of rat SC*

An approximately 15 week old Long-Evans rat was perfused intracardially with 4% paraformaldehyde (PFA). Following perfusion, the cervical enlargement of the SC was isolated. The animal experiment was performed in accordance with EU directive 2010/63/EU, and with approval from the animal ethics committee at the Champalimaud Centre for the Unknown (Lisbon, Portugal). Prior to imaging, the SC was washed in phosphate buffered saline (PBS). For imaging, the sample was positioned in a 5mm NMR tube and placed upright in a vertical bore 16.4T Bruker Aeon Ascend magnet with isotropic gradient strength of 3T/m (Bruker Biospin, Ettlingen, Germany). Diffusion weighted images were acquired axially with a pulsed gradient spin echo (PGSE) sequence with echo-planar readout and partial k-space coverage (70%) to accelerate acquisition. Image resolution was 36x36 $\mu m^2$ in-plane with a slice thickness of 1 mm. Diffusion encoding used the 9 directions needed for fast DKI (29,31). Nine non-zero b-values were acquired with nominal values linearly distributed between 0.6 ms/$\mu m^2$ and 5.4 ms/$\mu m^2$. In addition, 10 b=0 images were acquired for normalization. The remaining imaging parameters were TR/TE = 4000/35 ms, 4 averages, diffusion timings (Δ/δ) were 10ms/2ms. SNR > 60 at b=0. Sample temperature was maintained at 20 °C throughout the acquisition.

*3T MRI of human brain*

Human data was acquired in three normal volunteers and acquired with permission from the Ethics Committee of the Central Region, Denmark. Informed consent was obtained prior to data collection. Imaging was performed using a Siemens Trio 3T equipped with a 32 channel head coil. A twice refocused spin echo DW EPI sequence was used (diffusion time approximately 50 ms). Inversion recovery (inversion time of 2100 ms) was used to suppress cerebrospinal fluid (CSF) signal as per (40). Head motion during the acquisitions was avoided by inserting padding between the coil and the subject's head. The acquisition consisted of one b=0 image and 33 directions on 14 b-value shells between 0.2-3.0 ms/μm$^2$. The 33 direction sampling scheme was constructed by combination of a 3D 24 point spherical design (41) and the 9 directions identified previously (29). In all subjects, 19 consecutive slices were acquired at 2.5 mm isotropic resolution with TR=7200 ms, TE=116 ms, yielding an SNR of 39 at b=0.

*9.4T MRI of in vivo rat brain*

Data was acquired in four male Long Evans rats of equal age. Prior to imaging, each animal was anesthetized with isoflurane and placed in an animal cradle compatible with the rat brain cryocoil employed here. Once on the bed, anesthesia was maintained using isoflurane (1.5-2.5%) supplied through a nose cone. Animal respiration rate and temperature was monitored during the entire experiment. Animal temperature was maintained at 37°C using a heated water blanket. Positioning of the animal was done outside of the magnet using a transparent cryocoil replica ensuring that the animal is correctly positioned when inserted under the actual cryoprobe which is bore-mounted rendering direct animal positioning unfeasible. The experiment was approved by the Danish Animal Experiments Inspectorate (Dyreforsøgstilsynet permit no.: 2014-15-2934-01026). All animal handling was done in accordance with EU directive 2010/63/EU for animal experiments. Once the animal temperature and respiration rate had stabilized after positioning in the cradle, the animal was transferred to the scanner. Data was acquired using a Bruker Biospec 9.4T imaging system equipped with BGA-12HP gradients capable of 650 mT/m. We used a cross-coil setup with a 76 mm quadrature coil for excitation and a four element rat brain cryo-surface coil for reception. Gradients and all radio frequency coils were manufactured by Bruker Biospin (Ettlingen, Germany). The diffusion protocol consisted of a 1-9-9 acquisition using a segmented EPI sequence (four segments), diffusion times (Δ/δ) of 14ms/6ms, b-values 1.0 and 2.5 ms/µm$^2$; 3 b=0 images were acquired. Remaining scan parameters were: TE=27 ms, TR = 2237 ms, resolution 300 µm isotropic, 40 slices, matrix size 128x64, 20 averages were acquired resulting in a total scan time of 1h2m. SNR at b=0 was approximately 22.

*Postprocessing*

All data sets were denoised (42,43), Rice-floor adjusted (44), and corrected for Gibbs ringing effects (45). Following this, the data was inspected visually for quality (artifacts, subject/sample movement, and field drift). Due to careful padding around the head of each human subject, image registration was found to be unnecessary. Eddy current correction was performed in FSL (46), but was found to introduce an unsatisfactory amount of image movement, causing image registration to become necessary where it was not before. This correction step was therefore abandoned to avoid the blurring of the images that would be caused by motion correction. Image registration was also found to be unnecessary for the rat SC data due to the sample being tightly held in the NMR tube. The in vivo rat data showed very little movement due to the anesthetized rat being fastened to the animal bed for correct and stable placement under the cryocoil. This motion was corrected using linear registration to the b=0 volume in Matlab® (Natick, MA, USA).

Data analysis employed nonlinear least squares fitting performed with the Levenberg–Marquardt algorithm implemented in Matlab® (Natick, MA, USA). DKI analysis of the human data included only b-values up to 2.6 ms/µm$^2$. In the case of fitting to the axially symmetric DKI model, a good initial value for the axis of symmetry is crucial. For this, we used the primary eigenvector of D obtained from a diffusion tensor fit preceding the axisymmetrical DKI fit. Scripts for axisymmetrical DKI analysis are freely available from our group homepage (http://cfin.au.dk/cfinmindlab-labs-research-groups/neurophysics/software). The axially symmetric DKI representation was applied both to the full data sets, and to 1-9-9 subsets of the full data acquisition. For the rat SC data set, a 1-9-9 subset was extracted with b$_1$=1.3 ms/µm$^2$ and b$_2$=5.5 ms/µm$^2$ shells as in (27). For the human data, the 1-9-9 data set used b$_1$=1.0 ms/µm$^2$ and b$_2$=2.6 ms/µm$^2$ as per (31). All analysis steps are identical for aWMTI and faWMTI analysis. In this study, we adopt the criteria for inclusion of WM pixels used in (13) based on the Westin indices (47). Thus, for each data set a mask was generated based on the Westin indices of linearity ( $c_L = (\lambda_1 - \lambda_2)/\lambda_1 \geq 0.4$ ), planarity ( $c_P = (\lambda_2 - \lambda_3)/\lambda_1 \leq 0.2$ ), and sphericity ( $c_S = \lambda_3/\lambda_1 \leq 0.35$ ), where $\lambda_1, \lambda_2, \lambda_3$ are the eigenvalues of D in descending order. These indices were calculated from the diffusion tensor obtained from the conventional

DKI for all data sets except for the in vivo rat brain where the Westin indices where calculated from the axially symmetric diffusion tensor estimate from the axisymmetric DKI analysis. Our WMTI implementation is based on (13) but uses a slightly modified approach where the AWF is estimated simply as $AWF = K_{max}/(K_{max} + 3)$; this is exact when a direction with vanishing intra-axonal diffusivity exists – otherwise, this AWF is a lower bound estimate. This is less general than the full WMTI approach in (13) where a more involved alternative expression for the AWF is employed along with an optimization over the chosen WM region in which $D_a$ is then assumed to be constant. Such implementation choices may affect performance but are unlikely to affect the conclusions of the present study as our parameter estimates are generally in agreement with WMTI values in the literature (see also discussion). Conventional WMTI assumes the sign choice producing our Branch 1 (+ in Eq. (15c) and - in Eq. (15d)), which was shown to be stable over all directions in (13). Although this is less clear for Branch 2, we obtain our WMTI Branch 2 simply by using the opposite sign choice and assuming it to be stable over all directions. For the aWMTI and faWMTI the axisymmetric DKI parameters were processed with Eqs. (15) yielding the WMTI parameters directly.

For our analysis of the rat SC data we obtain a measure of the WM fiber dispersion in each pixel by fitting a model comprised of a Watson distributed collection of sticks and an extracellular compartment with all diffusivities as variables (48). From this fit the Watson concentration parameter $\kappa$ was obtained and used for calculating the average dispersion $\theta_C = \cos^{-1}\left(\sqrt{\langle \cos^2 \theta \rangle}\right)$ (relative to the out-of-plane direction) in each pixel, where angular brackets denotes the average over the Watson distribution:

$$\langle \cos^2 \theta \rangle = \frac{\int_0^\pi d\theta \sin\theta \cos^2\theta e^{\kappa \cos^2 \theta}}{\int_0^\pi d\theta \sin\theta e^{\kappa \cos^2 \theta}} \tag{16}$$

We truncate the intra-neurite signal expression after 12 terms, adequate for $\kappa < 128$.

*Numerical simulations*

One major difference between WMTI and the analytical WMTI variants is the assumption of perfect fiber alignment in the analytical framework (Eq. (15)). We therefore first compare the performance of WMTI and aWMTI using numerical simulations based on biexponential fitting to the human data detailed above. The biexponential model parameters were then used as ground truth values in our evaluation of the performance of the methods. For this evaluation, the diffusion tensor with the smallest minor eigenvalues was taken to belong to the intra-axonal space and its volume fraction was used as the true AWF. In the simulations, the experimental signals from 100 random WM pixels satisfying the Westin index criteria were fit to the biexponential signal equation (Eq. (2)) with non-coaxial, unconstrained diffusion tensors (i.e. all entries on the diagonal of $D_a$ are allowed to be non-zero to provide the most flexible fit and to account for effects of intra-voxel fiber dispersion in the Gaussian limit. Here, all shells over the acquired b-value range (0-3.0 ms/µm$^2$) were employed. Non-linear least squares fitting was performed using the Trust Region Reflective algorithm in Matlab (Natick, MA, USA). These fits reveal a typical separation of several degrees between the primary eigenvectors of the two tensors. The parameter values obtained from these fits were then used with the biexponential model to generate synthetic data sets using the same encoding scheme as the human experiments and a maximum b-value of 2.6 ms/µm$^2$. Noise was added in quadrature to an SNR matching the experimental of 39 (i.e. the simulations do not take into account the denoising applied to the experimental data). A total of 1000 noise realizations was performed in each of the 100 pixels. The generated signals were then analyzed in the same manner as the experimental data to yield WMTI and aWMTI parameters.

## Results

Figure 1 shows the results (histograms of relative errors) of the numerical simulations using the biexponential model parameters as ground truth for each pixel. The figure text reports median and mid-95% range of the error for each parameter and each of the WMTI/aWMTI estimation strategies over 100 WM voxels and 1000 noise realizations in each voxel. Here, only the branch yielding estimates in agreement with the true values are shown (Branch 1 in all cases). Input parameters (volume fraction, summary parameters of the tensors from the 'fast' and 'slow' diffusion components) from the biexponential fits to the 100 random WM pixels are shown in Supplementary Fig. 1. We ensure the relevance of our input parameters by comparing to values obtained from the high quality data from human brain provided in (37). Overall, good agreement is seen between our input values and the reference values derived from data acquired with sampling to high b-values. Nevertheless, some differences are seen which likely stem from the reference data being produced by a fit to the average signal in a WM region whereas values employed in our simulations stem from 100 random WM pixels, as well as differences in the employed b-value range.

Figure 1 shows that both methods perform quite well and that they generally agree. Closer scrutiny reveals aWMTI to have lower median error than WMTI for all parameters except $D_a$ and tortuosity (α). However, also here the methods show very similar performance. The error range is also quite similar for the two methods, with only the $D_a$ estimate showing markedly different behavior between the two with aWMTI having the largest error range.

Turning to analysis of the experimental data we first compare the performance of WMTI to aWMTI in the WM fibers in rat SC (3551 voxels satisfying the Westin index based criteria detailed above). The same five parameters as in Fig. 1 are compared in Fig. 2 which includes estimates for both sign choices (Branch 1 and Branch 2). For all parameters, aWMTI estimates are seen to correlate strongly to their WMTI counterparts. For Branch 1, very strong correlations are seen with most estimates clustered tightly around the identity line (red). AWF displays the weakest linear correlation of the six parameters with a correlation coefficient of 0.9. Here, as well as in the rest of the study, all reported correlations are Pearson's linear correlation coefficients significant at p<0.05. Estimates in Branch 2 also show very strong correlation, but here larger offsets from the identity are observed. Thus, the assumption of axial symmetry does not negatively affect WMTI estimation in this tissue. It is important to stress that the assumption of axial symmetry does not mean that the aWMTI and faWMTI methods assume or require perfect alignment of fibers in the tissue. To illustrate this, Fig. 3A shows the average fiber dispersion ($\theta_C$) in the SC in each pixel. The red outline shows the WM region in the rat SC analyzed throughout. The histogram in Fig. 3B shows the distribution of $\theta_C$ in this sample with values varying from 26° up to 47° (the WM average is 38°). WM $\kappa$ values range from 1.5 to 5.5 with an average value of 3.6. When comparing aWMTI to faWMTI in the SC (data not shown) the correlations unsurprisingly decrease but remain very strong (all exceed 0.82). Maps of all parameters from WMTI, aWMTI and faWMTI in the SC are provided as Supplementary Figs. 2-4. Visual inspection of the AWF maps from rat SC reveals the expected left right symmetry and regional variation of AWF seems to correspond to known WM tract locations as segmented in (49).

Figure 4 shows the same type of analysis as in Fig. 2, but this time for suitable WM regions in one normal human brain (4401 voxels across all slices). Here, aWMTI estimates in Branch 1 correlate very strongly with WMTI estimates with all correlations exceeding 0.86. Interestingly, the behavior in Branch 2 is different (as was also seen in Fig. 2) where correlations for $D_{e,\|}$ and tortuosity are very strong but the $D_a$ correlation is only moderately strong at 0.68. Similar behavior is observed when comparing aWMTI to faWMTI in this subject. Here, correlations in Branch 1 are in the range of 0.67-0.74 except for the tortuosity where propagation of noise from the faWMTI estimates of $D_{e,\|}$ and $D_{e,\perp}$ causes the correlation to decrease to 0.61. In Branch 2 the correlation for $D_{e,\|}$ is 0.72 whereas $D_a$ and tortuosity correlations drop markedly to 0.44 for $D_a$ and 0.57 for $\alpha$. Average correlations for all three subjects (aWMTI vs WMTI, aWMTI vs faWMTI) are given in Table 1.

Figure 5 shows an example of faWMTI estimates (in suitable pixels) of AWF, $D_{e,\perp}$ and $\alpha$ overlaid on the $S_0$ image in normal rat brain at 300 μm isotropic resolution. Two orthogonal slice planes are shown. The parameter estimates lie in the expected range, vary smoothly, and display the expected left-right symmetry.

*Assessment of biophysical parameter values*

In Fig. 6 we show histograms of WMTI estimates of $D_{e,\|}$ and $D_a$ for both branches (Branch 1 in top row, Branch 2 in bottom row) in suitable pixels in rat SC (column A) and normal human brain (column B). Column C shows faWMTI estimates from in vivo rat brain. Columns B and C show pooled values from all subjects/animals. Analysis of each subject/animal separately showed the same overall behavior. The free water diffusivity at the sample temperature is marked with a vertical red line in all graphs as it provides an upper bound for credible parameter values. The difference in ranges between the left column and the other two columns is due to the sample temperature: approximately 20 °C (where free water diffusivity is $\approx 2$ μm$^2$/ms) for the rat SC (column A), and 37 °C (where free water diffusivity is $\approx 3$ μm$^2$/ms) for the human and rat brain (columns B-C). For all three systems, Branch 1 produces diffusivity values below the upper bound imposed by free water diffusivity except for a tail of high $D_{e,\|}$ values in the rat brain (column C, top panel). In the case of Branch 2, the $D_{e,\|}$ values are also physically plausible, but the vast majority of estimated $D_a$ values exceed free water diffusivity. Fig. 7 shows the correlation between $\theta_C$ and WMTI estimates of $D_a$ and $D_{e,\|}$ for both branches in rat SC WM. The black line shows a robust fit to the data. Branch 1 is seen to display the expected behavior of decreasing diffusivity for increasing dispersion whereas Branch 2 does not.

We note a significant negative offset between the aWMTI and WMTI branch 2 estimates (Figs. 2 and 4) although they correlate strongly. Since our WMTI Branch 2 assumes sign stability which we cannot at present verify, we also present aWMTI Branch behavior (Fig. 8). Here, branch behavior from SC and human brain (Fig. 8A-B) is shown with Branch 1 (2) estimates shown in the top (bottom) row. Interestingly, aWMTI Branch 2 estimates of diffusivities violate the physical upper bound in fewer pixels than WMTI (Fig. 7), albeit still in approximately 50% of WMTI pixels in human brain (Fig. 8B). In SC both branches largely produce diffusivities within the physical bounds with 83% of Branch 2 $D_a$ values below the upper bound of 2 µm$^2$/ms (Fig. 8A). Correlations between diffusivities from both aWMTI branches in SC and WM fiber dispersion are shown in Fig. 8C. We note that both branches now display the expected decreasing diffusivity with increasing dispersion. This behavior, however, remains most pronounced in Branch 1 as indicated by the correlations above each plot.

## Discussion

This work extends earlier work on WMTI by combining a time-efficient data acquisition strategy from our earlier work (27,29-31) with a post-processing strategy utilizing analytical relations between WMTI parameters and tensor components similar to the originally proposed WMTI method which was derived assuming perfectly aligned WM bundles (28). The current standard for WMTI, however, builds on later work where the assumption of perfect alignment was relaxed to allow for angular spread up to 30° (13). This allows WMTI to be performed in most of the brain's major WM tracts, where WM is highly aligned yet shows some dispersion. For instance, histology and N-acetylaspartate (NAA) diffusion spectroscopy show fiber dispersion in the human corpus callosum to be significant (group average 18°) (50) in agreement with similar histological analysis in the rat (51). Similarly, our findings in rat SC show a significant degree of dispersion in WM (see below).

In order to perform WMTI based on scarce data such as the 1-9-9 protocol for fast kurtosis (29-31) a reduction of parameters in the DKI signal representation is needed. Here, this is achieved by assuming axially symmetric diffusion and kurtosis tensors as in (27). However, as detailed above, axial symmetry is not completely fulfilled in even very aligned WM bundles, and is not assumed in conventional WMTI. We therefore performed simulations to evaluate the performance of aWMTI against WMTI and ground truth parameters. This was done using the biexponential signal model (Eq. (2)) to synthesize realistic non-axisymmetric DKI signals which are then analyzed with WMTI and aWMTI. Supplementary Fig. 1 shows our simulation input parameters to agree well with those obtained from high quality data acquired up to high b-value in human WM (37). The simulations (Fig. 1) show that aWMTI extracts the ground truth simulation parameter values with less bias but more spread than WMTI in most cases. However, overall the performance of the two methods is highly comparable meaning that the assumption of axially symmetric diffusion and kurtosis tensors does not impede aWMTI estimate fidelity in WM compared to WMTI. The presented results are based on direct estimation of parameters using Eqs. (15). An alternative strategy using least squares estimation with Eq. (12) was also evaluated with very similar performance to aWMTI in Fig. 1 was seen with this approach (data not shown). In our remaining analysis, we continue to compare aWMTI to WMTI for consistency with the literature where WMTI is the standard method for which validation studies have been carried out. As mentioned in the Methods section our WMTI implementation is slightly simplified compared to the full WMTI framework in (13). While our parameter estimates are in agreement with the literature (all report only Branch 1) we do stress that implementation details such as fitting strategies may affect performance (52,53). Moreover, the assumption that sign choice is stable over all directions may not be true for Branch 2. Analysis of whether this assumption produces a proper WMTI Branch 2 implementation is an interesting (non-trivial) topic for future work.

Very strong agreement between WMTI and aWMTI is also seen in Fig. 2 which shows scatter plots based on rat SC data. Here, correlations in both branches all exceed 0.93 except for AWF (0.9). Even when a 1-9-9 data subset is used for faWMTI, correlations to aWMTI remain strong (>0.83) in both branches. Figure 3A illustrates SC WM fiber dispersion by showing the average fiber dispersion, $\theta_C$, in each pixel. The average across the WM is 38°. Stating these results in terms of the Watson concentration parameter $\kappa$, we find $\kappa$ = 3.6 on average in WM. To put this into perspective, for a fiber arrangement characterized by a Watson distribution with $\kappa$ =4, only 20% of the fibers have angles ≤15° relative to the main direction, and a dispersion range as wide as 60° is needed to account for 91% of the fibers (54). Perfect alignment is thus far from fulfilled even in the SC. We note, that these results might be expected for an acquisition with a slice thickness of 1 mm in a section of the cervical SC where many nerve branches exit the SC to the extremities. Thus, the SC analysis shows that the aWMTI method is capable of producing robust estimates even in geometries where perfect alignment is not fulfilled which is in agreement with our simulations. Visual inspection of the parameter maps from rat SC (in Supplementary Figs. 2-4), shows regional variation of e.g. AWF which might indicate ability to map individual WM tracts in SC with WMTI methods, but more samples and histology would be needed to verify this. We note that fixed rat SC seems very well-suited for future validation studies of the WMTI methods with histological analysis of various WM tracts as in (49).

Turning to the human data, correlations between WMTI and aWMTI are also strong (Fig. 4) but slightly lower than those seen in fixed tissue (Fig. 2). This is most likely due to the lower SNR and presence of physiological noise in the human data. Nevertheless, the overall behavior in the method comparisons in Figs. 2 and 4 is very similar except for the Branch 2 estimate of $D_a$ in human brain, where a correlation of 0.68 is seen compared to 0.94 for rat SC. In both cases, Branch 2 correlations fall far from, and below, the identity line. The Branch 1 estimates all agree with the value ranges and distributions presented from normal human brain in (13).

As expected, the correlations decrease when reducing the data foundation to a 1-9-9 subset of the human data. Nevertheless, the correlations remain strong between faWMTI and aWMTI (average correlations >0.7 for AWF, $D_a$, and $D_{e,\perp}$, >0.6 in remaining cases for Branch 1); see Table 1 for full results. Interestingly, the estimation fidelity is not the same in the two branches with $D_a$ correlations in Branch 2 being much poorer than in Branch 1. Branch 2 estimates therefore seem more sensitive to the reduction in data (faWMTI) than estimates in Branch 1.

The proposed framework provides a means of reducing the required number of images. Since scan time can be a constraint in most clinical settings, such a technique may be useful. When scan time is less of a concern, the lower data demand can be utilized to achieve higher data quality, i.e. higher resolution and/or SNR. Higher resolution data may increase the number of WM pixels with a uniform fiber orientation thus further improving the agreement with aWMTI assumptions. However, it is a subject for future investigations to determine whether high SNR or higher angular sampling is optimal for WMTI. A fast alternative to conventional WMTI may therefore be useful in the clinic and for clinical and preclinical research including validation studies where both high image resolution and whole brain coverage are desirable but perhaps not feasible with WMTI based on a conventional DKI acquisition. An example pointing to the usefulness of the faWMTI method for preclinical research is given in Fig. 5, which shows faWMTI mapping of AWF, and $D_{e,\perp}$, and tortuosity (Branch 1 throughout) in rat brain at an unprecedented resolution of 300 μm isotropic resolution. Given the degree of correlation values between aWMTI and faWMTI in human brain, some noise in the estimates might be expected, but they are seen to vary smoothly and display the expected symmetry between hemispheres. Similar behavior is seen in all four rats.

*Validation and biophysical parameter duality:*

WMTI has been shown to provide valuable WM biomarkers in several validation studies demonstrating correlation between WMTI and tissue parameters derived from histology or measured with electron microscopy. Experimental validation was offered in (13), and more recently in (21-23). In the three latter studies, the cuprizone model of demyelination (55-57) was used. In (23) histology was used for validation of DKI based WM modeling. Here, AWF, mean kurtosis (MK) (2) and radial kurtosis (RK) (2,26) were found to be the most sensitive markers for the cuprizone induced WM changes. Similar findings were reported in (21) with MK, RK and AWF deemed the most sensitive DKI and WMTI parameters for detection of cuprizone induced WM changes in corpus callosum. Overall, a range of DTI, DKI and WMTI parameters were found to discriminate the cuprizone and control groups in various brain regions and in different stages from acute to long lasting changes. High resolution WMTI was performed in ex vivo mouse brains from knock-out models showing varying degrees of hypomyelination in (24). As in (21), the authors conclude that DKI offers improved sensitivity over DTI to myelination changes and exhibit stronger correlation to myelin from histology than DTI metrics. The authors also conclude that AWF from WMTI is a reasonably accurate reporter of axon water fraction in near normal WM compared to estimates from histology. AWF from WMTI was found to correlate significantly with total AWF derived from electron microscopy (EM) in (22). In that same study, $D_{e,\perp}$ was found to correlate with the WM g-ratio (the ratio between the axon diameter alone to the diameter of the myelinated fiber) from EM but not with the AWF from EM. These parameters (AWF and $D_{e,\perp}$), are therefore strong candidates for MR-derived markers with specificity to demyelination and remyelination. It is important to note that the estimates of AWF and $D_{e,\perp}$ are unaffected by branch choice. Other parameters ($D_{e,\parallel}$, $D_a$, and $\alpha$) are, however, strongly affected by the choice of sign as also shown throughout our analysis. Typically, this has been resolved by a sign convention in WMTI yielding solutions such that $D_a \leq D_{e,\parallel}$ (13). However, this has recently become a topic of debate e.g. in (15,32,38), where in the latter reference it is shown that arguments can be made in favor of either $D_a \leq D_{e,\parallel}$ or the opposite. Figure 6 summarizes the observed branch behavior for WMTI estimates of $D_{e,\parallel}$ and $D_a$ in all three systems employed here (column A: rat SC, column B: in vivo human brain, column C shows faWMTI estimates from in vivo rat brain). Overall, our analysis shows WMTI Branch 2 to produce $D_a$ estimates in excess of the free water diffusivity (vertical red line) in a substantial number of pixels. Branch 1 estimates are generally within the physical range. Figure 7 further points to the physically reasonable behavior in Branch 1 where both $D_a$ and $D_{e,\parallel}$ are seen to decrease with increasing fiber dispersion in rat SC WM. Branch 2 does not display this behavior. A similar analysis correlating AWF to $D_a$ and $D_{e,\parallel}$ was inconclusive (not shown).

Our results are generally in agreement with previous WMTI literature where the choice leading to $D_a \leq D_{e,\parallel}$ (Branch 1) has typically been favored (13,28). However, in live rat we see a large overlap of diffusivity estimates within the physically acceptable range in both branches (Fig. 6C). Interestingly, the Branch 2 $D_a$ estimate has a second peak at approximately 1.7 µm²/ms, which agrees with the overall $D_\parallel$ of water measured in vivo in rat SC (58). We also note that our in vivo rat data do show some unphysical behavior of $D_{e,\parallel}$ in Branch 1 (Fig. 6C top panel). Comparing the histogram of error in WMTI $D_a$ estimation in our simulations (Fig. 1) to the Branch 2 $D_a$ estimate in rat SC in Fig. 6A the spread around the true value in Fig. 1 is seen to be very similar to the distribution around the free diffusion value in SC. This might indicate that the Branch 2 $D_a$ in SC is close to the free water diffusivity of 2 µm²/ms. In this case neither branch can be rejected based on the diffusivity estimates.

This notion is further supported by the histograms of aWMTI branch estimates (Fig. 8A-B) which shows aWMTI branches to have a somewhat different range than the diffusivities in the WMTI branches (as also seen in Figs. 2 and 4). In particular, the SC Branch 2 behavior is now seen to be mostly within the physically plausible range (Fig. 8A) and both show decreasing diffusivity with increasing dispersion (Fig. 8C). This is worth noting because our WMTI Branch 2 implementation is based on the assumption that the non-conventional sign choice is robust over directions as is the case for the conventional branch (Branch 1). This assumption may not be correct and the lower values produced by aWMTI in Branch 2 therefore cannot be ruled out as mere bias. We note that the Branch 2 aWMTI $D_a$ estimate in human brain is centered on the free water diffusion coefficient (Fig. 8B) as discussed for the SC. Clearly, there is a difference in the results obtained in fixed tissue and in vivo. It is unclear if estimation uncertainty in the presence of physiological noise could explain the in vivo Branch 2 $D_a$ exceeding the free water value to the extent seen here. Nevertheless, our analysis seems to indicate that if the intra-axonal diffusivity is almost free, estimation uncertainty might be the cause of some apparent unphysical behavior. In this case neither branch can be ruled out. This is further underscored by the result in Fig. 8C, where both aWMTI branches display the expected negative correlation between dispersion and diffusivity. This behavior, however, is still most pronounced for Branch 1 as in Fig. 7.

Although our analysis cannot resolve the debate over the correct branch, our analysis does fall in line with the literature in that it mostly favors the conventionally chosen Branch 1. However, as pointed out above our results do contain some ambiguities in agreement with recent developments and observations either favoring Branch 2 (38,59) or even suggesting $D_a \approx D_{e,\parallel}$ in rat SC in vivo (58). It should also be pointed out that our results - both in our analysis as well as in our simulations - may be determined by the data foundation, i.e. that we are bound to obtain faulty Branch 2 behavior due to the manner in which our data is acquired (essentially forcing all optimizations down one of the 'pipes' described in (32)). Higher b.-value acquisitions and advanced analysis frameworks as the one proposed in (38) may resolve the ambiguity and will likely aid in optimizing experimental procedures. We note, that a data set similar to the human brain data sets used here is publicly available for those who wish to use a similar data foundation to compare our results to results from other analysis methods (60). Data sharing may be valuable in further investigation of the branch behavior as both data acquisition details and analysis scheme may affect which branch yields physically plausible estimates (38). We note, that in our analysis the branch estimates also seem to respond differently to the reduced data amount in faWMTI indicating different noise sensitivity in the two branches. Besides advances in analysis, experiments to resolve the duality problem may be possible, e.g. by investigation of the time dependence of parameters in both branches, or by direct experimental observation. In (61), the apparent diffusivity of water was mapped in the soma and initial segment of the axon in intact neuron, in situ. Such measurements using cellular level MR microscopy (61-65) may aid in resolving this modeling ambiguity by providing estimates of diffusivities in specific tissue compartments. So far, only fixed tissue has been examined in this manner but future experiments may be possible where the perfused acute brain slice model can be employed in a microscopy setup as in (66).

Caution is needed when comparing results between such different systems as employed here, and with somewhat varying experimental procedures. Our human data was acquired with CSF suppression as recommended in (40), but this was not employed in the remaining acquisitions. A post-hoc analysis of a human data set acquired without the inversion recovery preparation showed the same estimate behavior between branches (data not shown) indicating that CSF suppression does not affect the WMTI estimates much. Similarly, the overall branch behavior was the same in an analysis omitting the preprocessing steps employed here. Such details therefore do not seem to be responsible for the observed branch results. Other differences between data sets include biophysical differences between the fixed and in vivo state (as mentioned above), sequence details, and experimental field strength (affecting relaxation properties which may vary between compartments (67)). Since the echo times employed in the acquisitions are also very different (particularly between the preclinical data and the human data) compartmental differences in transverse relaxation may also contribute to differences observed between the systems. Most likely, the primary difference however, lies in the applied diffusion timings where for the human data, the diffusion time of approximately 50 ms is long enough to ensure that the Gaussian fixed-point asymptote is reached (i.e. no compartmental kurtosis survives at these times), as is assumed in the WMTI framework. However, in the rat SC and in vivo rat brain, diffusion times are shorter and the tortuosity regime may not be fully reached, potentially causing a mix of contributions to overall kurtosis (different compartmental diffusivities and kurtosis) to be captured in the measurements. The diffusion time dependence of the DW signal was investigated in rat cortex in (68). Here apparent kurtosis was seen to sharply increase from the lowest diffusion times of a few ms up to approximately 10 ms where it peaked and showed a slight decrease (from 0.60±0.05 to 0.51±0.05 , values read from figure in (68)) with increasing diffusion time (measured up to 30 ms). Their analysis also showed a negligible diffusivity variation in this time range. Assuming that intra-cellular kurtosis had vanished at the longest diffusion time we can estimate that the intracompartmental kurtosis contributes roughly 15% of the peak kurtosis observed at 10 ms. Although these considerations stem from observations in gray matter, our rat SC data (Δ = 10 ms) and rat brain data (Δ = 14 ms) may be somewhat similarly affected, particularly the SC data as it was acquired under conditions where diffusion is slower (fixed tissue at room temperature) than in vivo. More experiments are needed to elucidate these matters in WM. Since the data acquisition details are the same in our comparison of WMTI, aWMTI and faWMTI the results of the main topic of this study - characterization of WMTI based on axially symmetric DKI - are unaffected. Validation and correct estimation and interpretation of the biophysical parameters are, however, highly important problems, as are the effects of time-dependence. With the faWMTI method, acquisition of data sets spanning a range of diffusion times becomes more experimentally feasible than with previous approaches. Investigations of the diffusion time dependence of WMTI parameters are ongoing in our group.

In addition to WMTI other novel WM markers may be of value. One example is the kurtosis fractional anisotropy (KFA) (29,69), which has been shown to offer WM contrast where FA does not (39,70). Post-hoc analysis (data not shown) shows that estimation of KFA is feasible with axisymmetric DKI, but the agreement is best for high SNR data such as the rat SC data. In human brain, KFA estimated with axisymmetric DKI correlates strongly (>0.9) with KFA from conventional DKI but a marked loss of contrast in KFA from axisymmetric DKI hints that information is lost by imposing axial symmetry on the tensors. The KFA analysis also shows that, unlike the central DKI and WMTI parameters, the 1-9-9 protocol is not adequate for estimation of KFA in whole brain where KFA contrast further deteriorates. This is in agreement with the results in (39), where a rapidly obtainable KFA proxy based on the 1-9-9 protocol was investigated, but found unfeasible due to high SNR requirements.

## Conclusion

We presented and evaluated WMTI based on a simplification of the DKI signal expression obtained by imposing axial symmetry on both tensors D and W and an analytical framework. The performance of this strategy and the effect of the imposed tensor symmetries on WMTI parameter estimation in non-axisymmetric systems was evaluated using numerical simulations. In general, the proposed approaches display improved or similar performance over conventional WMTI estimates when compared to simulation ground truth values. Correlations were then investigated between WMTI and axisymmetric WMTI estimates based on large data sets (aWMTI) and sparse data sets obtained with the 1-9-9 protocol for fast kurtosis estimation (faWMTI). In the analysis of experimental data from fixed rat SC and human brain, very good agreement was seen between WMTI and aWMTI parameter estimates in most cases. Reducing the data foundation to a 1-9-9 acquisition caused the correlations to decrease, but strong correlations between aWMTI and faWMTI persisted for all of the parameters - importantly also for parameters that have shown potential as WM markers in validation studies. Lastly, we presented in vivo faWMTI in rat brain with isotropic resolution of 300 µm, demonstrating the preclinical potential of the method. We provided an overview of parameter estimates from both branches of a solution ambiguity across all investigated systems. Although a number of potential confounds exist, overall, our analysis indicates that the conventionally chosen branch (Branch 1 where $D_a \leq D_{e,\|}$) most consistently leads to physically plausible predictions. While not conclusive on the matter of appropriate branch choice, our aWMTI/faWMTI methods may contribute to the current debate over WMTI parameter estimation by providing a technique for efficient data acquisition for investigation of e.g. parameter time dependence and for high resolution validation studies. Furthermore, the proposed faMWTI technique based on the fast kurtosis strategy opens interesting clinical possibilities where now most DKI techniques can be explored and applied in routine clinical MRI even in very demanding workflows.

## Acknowledgments

The authors were supported by the Danish Ministry of Science, Technology and Innovation's University Investment Grant (MINDLab, Grant no. 0601-01354B). The authors acknowledge support from NIH 1R01EB012874-01 (BH), Lundbeck Foundation R83-A7548 (SNJ). The 9.4T lab was funded by the Danish Research Council's Infrastructure program, the Velux Foundations, and the Department of Clinical Medicine, AU. The 3T Magnetom Tim Trio was funded by a grant from the Danish Agency for Science, Technology and Innovation.

## Tables and figure captions

|  | Branch 1 | | | | | Branch 2 | | |
| --- | --- | --- | --- | --- | --- | --- | --- | --- |
|  | AWF | $D_a$ | $D_{e,\parallel}$ | $D_{e,\perp}$ | $\alpha$ | $D_a$ | $D_{e,\parallel}$ | $\alpha$ |
| aWMTI vs WMTI | 0.86±0.01 | 0.84±0.05 | 0.95±0.01 | 0.96±0.01 | 0.95±0.01 | 0.67±0.02 | 0.77±0.08 | 0.90±0.01 |
| aWMTI vs faWMTI | 0.73±0.02 | 0.75±0.01 | 0.66±0.03 | 0.70±0.02 | 0.61±0.02 | 0.47±0.03 | 0.72±0.01 | 0.59±0.02 |

Table 1: Summary of results in human brain when comparing aWMTI to WMTI, and aWMTI to faWMTI. Values are average correlations ± one standard deviation across three human subjects for all parameters (both branches).

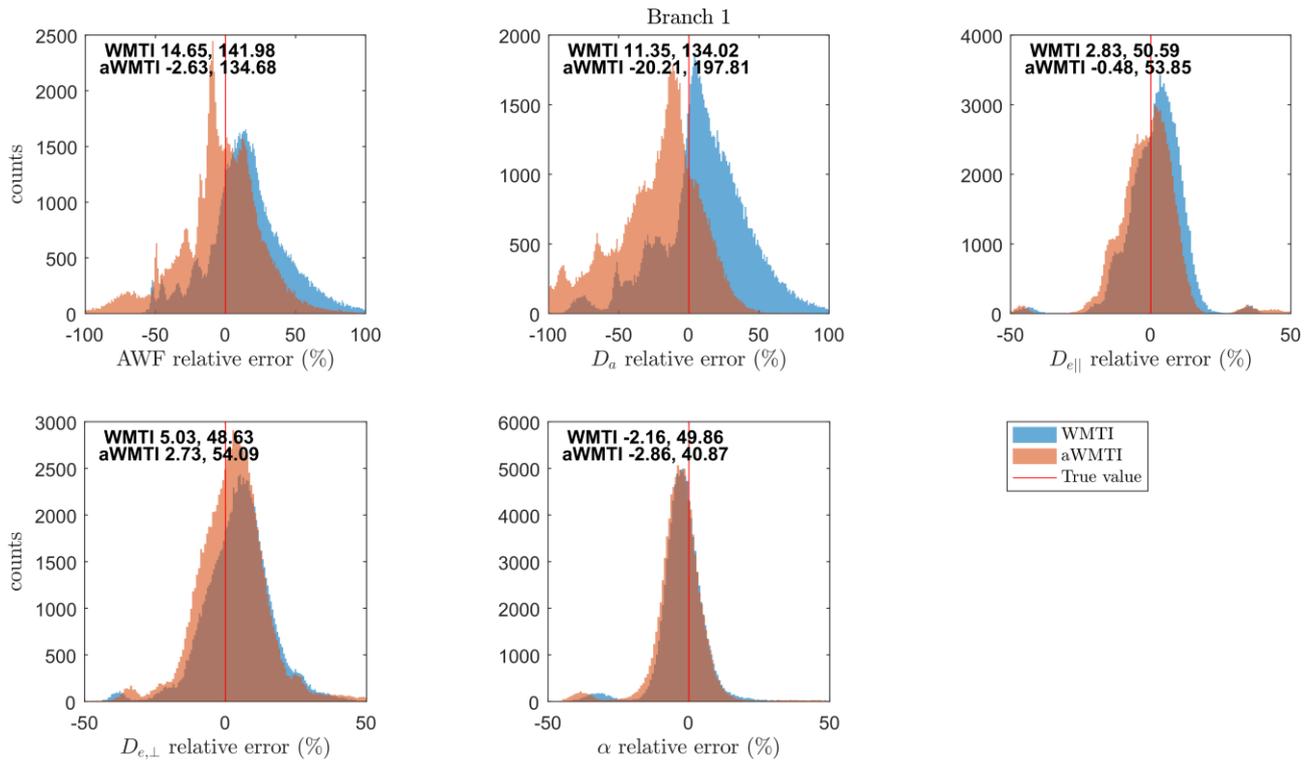

Fig. 1: Results from numerical simulations comparing performance of WMTI to aWMTI. Histograms show relative errors compared to ground truth values for each method. The red line marks zero error. Text inside each plot reports median relative error and mid-95% range for each method.

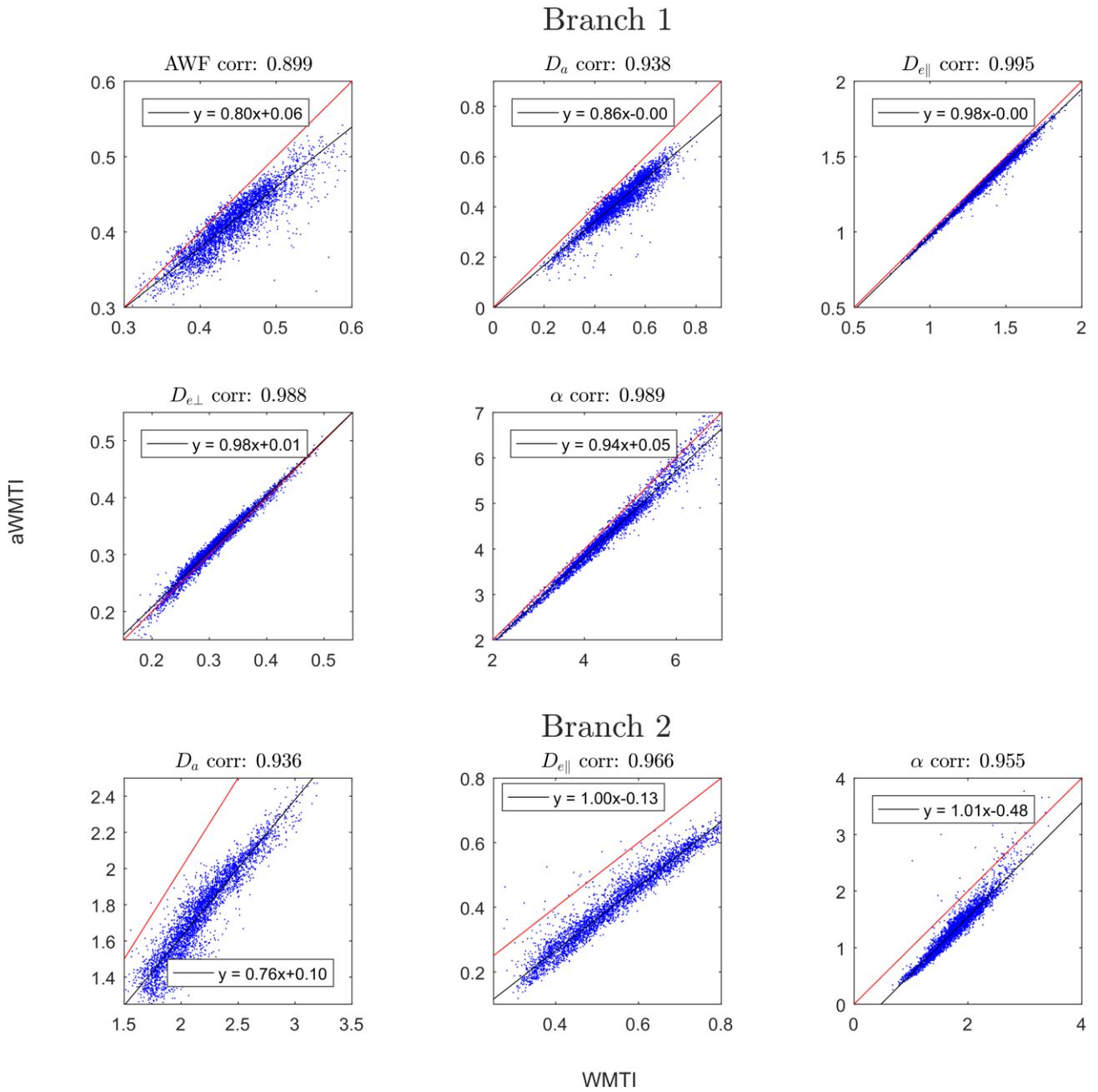

Fig. 2: Comparison of parameter estimates from WMTI and aWMTI in rat SC white matter.

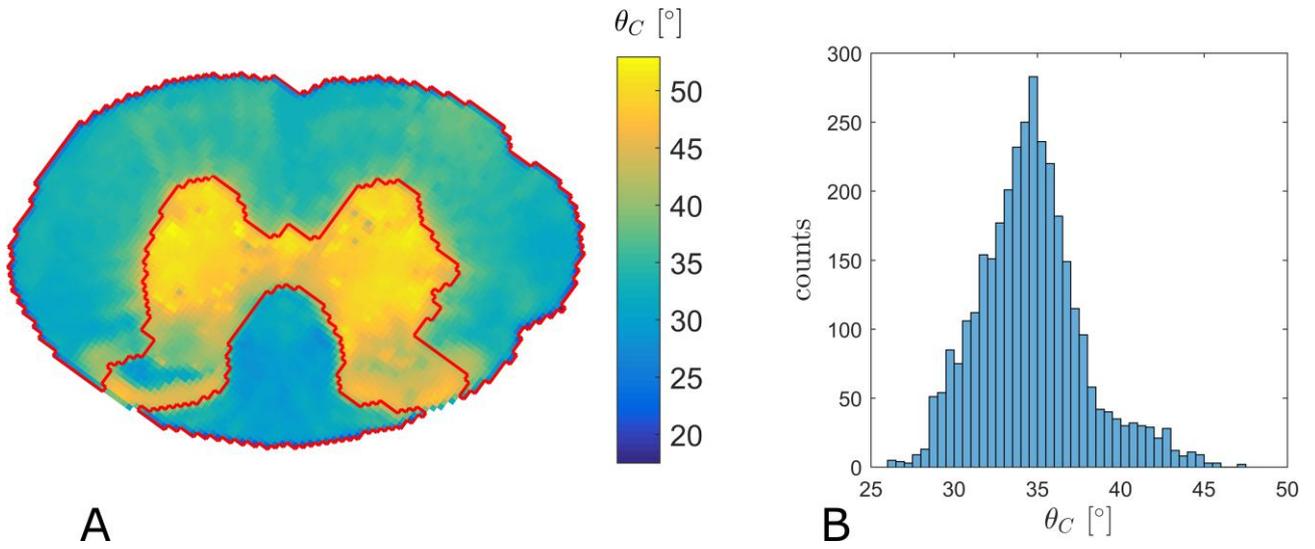

Fig 3: A) Mapping of the average intra-voxel fiber dispersion (in degrees) in the rat SC. The red outline shows the WM region analyzed throughout. B) Histogram of the average intra-voxel fiber dispersion mapped in panel A.

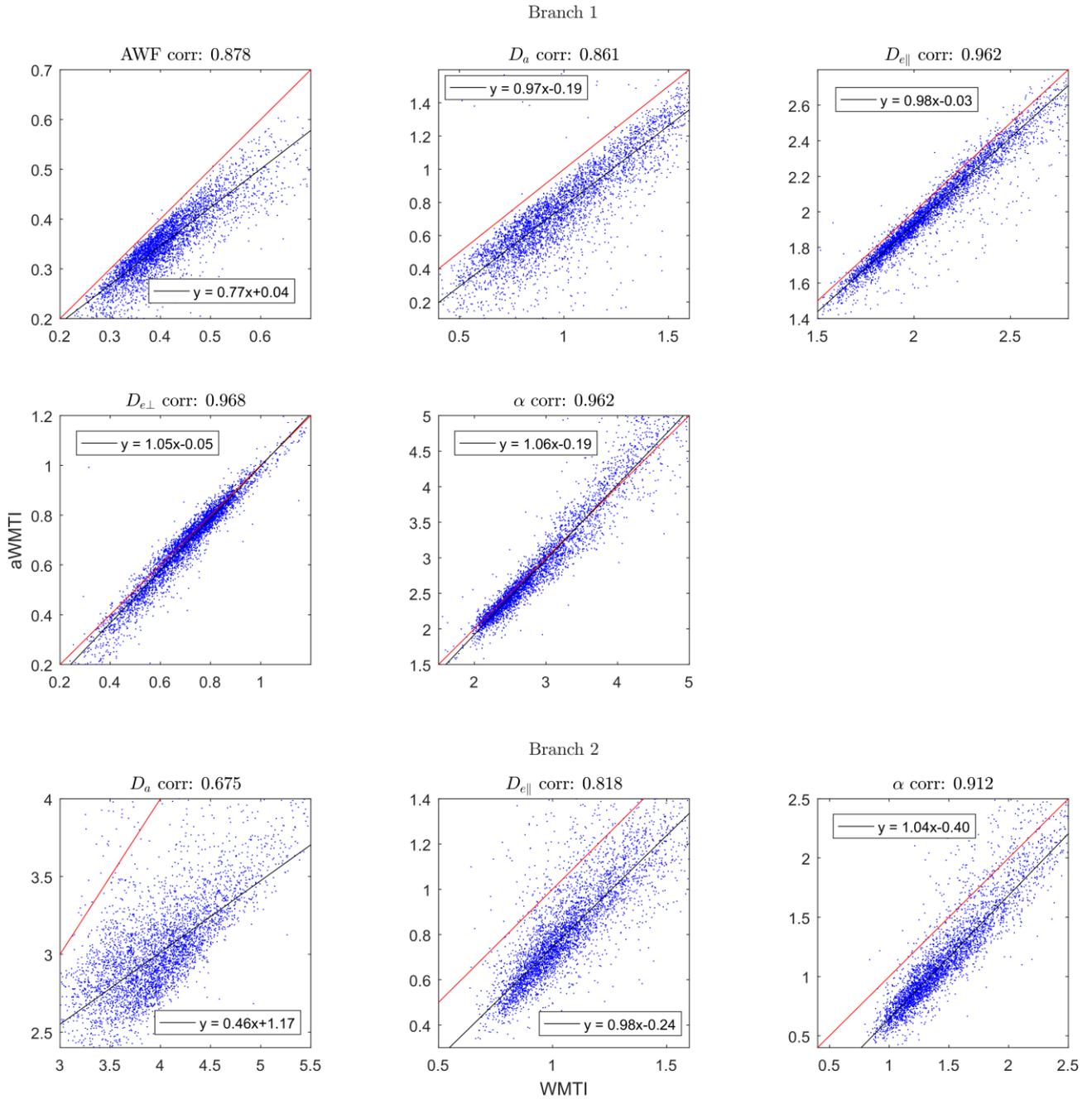

Fig. 4: Scatterplots comparing WMTI and aWMTI estimates in both branches in whole brain (4401 WM pixels) of one human subject. On average less than 9% of the data fall outside the shown ranges.

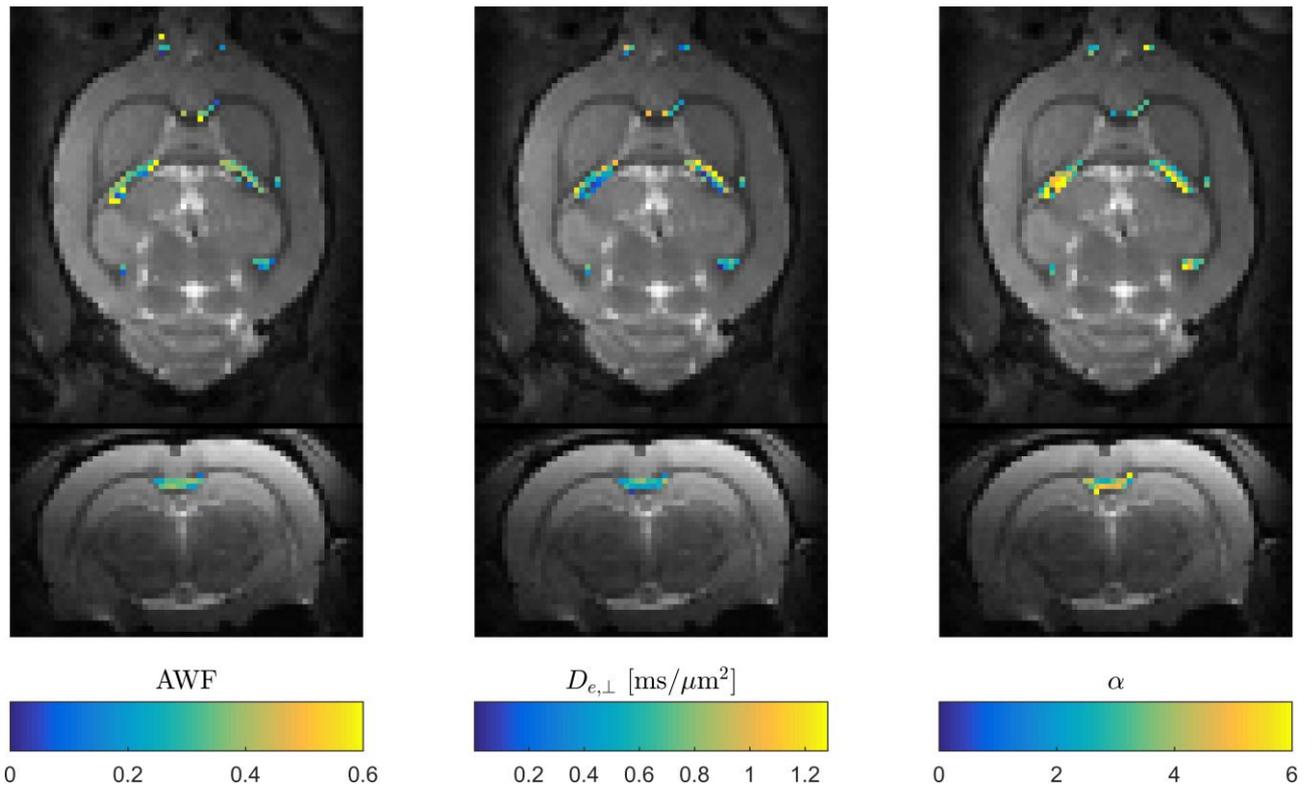

Fig 5: Data example showing estimates of three faWMTI parameters (Branch 1) based on a 1-9-9 acquisition in live rat brain at an isotropic resolution of 300 μm. Axial and coronal slice planes are shown. The parameter values are overlaid on the b=0 images from the 1-9-9 data set.

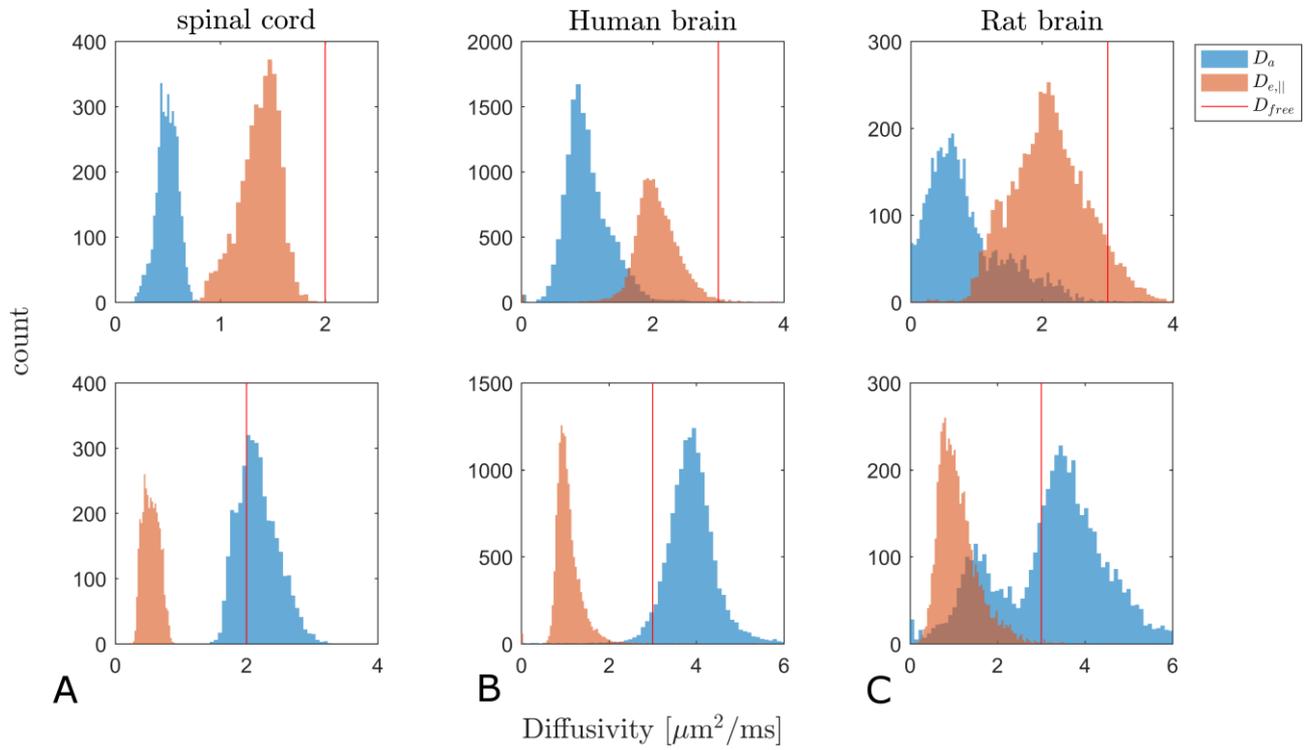

Fig. 6: Histograms of WMTI estimates of $D_{e,\|}$ and $D_a$ for Branch 1 (top row) and Branch 2 (bottom row) in rat SC (column A), human brain (column B). Column C shows faWMTI estimates from in vivo rat brain. Columns B and C show data from all subjects/animals. Vertical red lines mark the free water diffusivity at the sample temperature.

# Branch 1

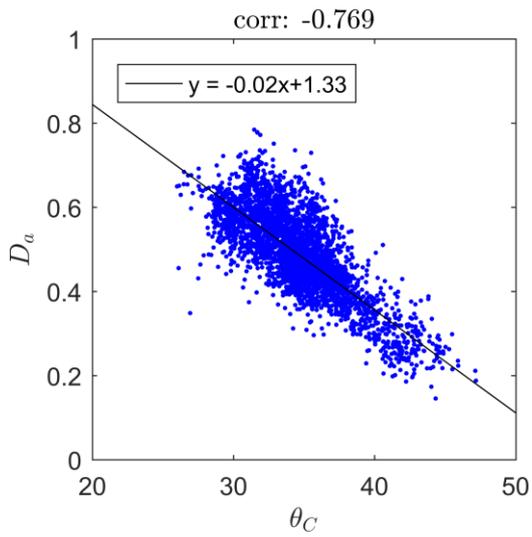
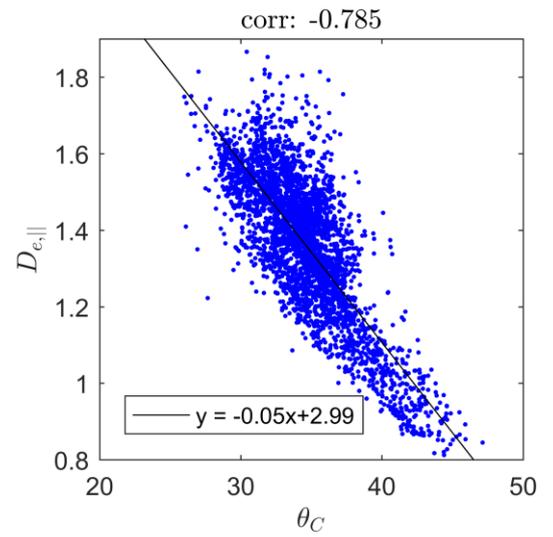

# Branch 2

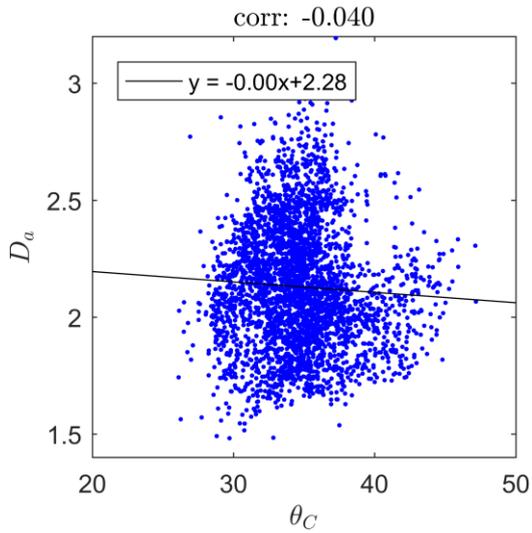
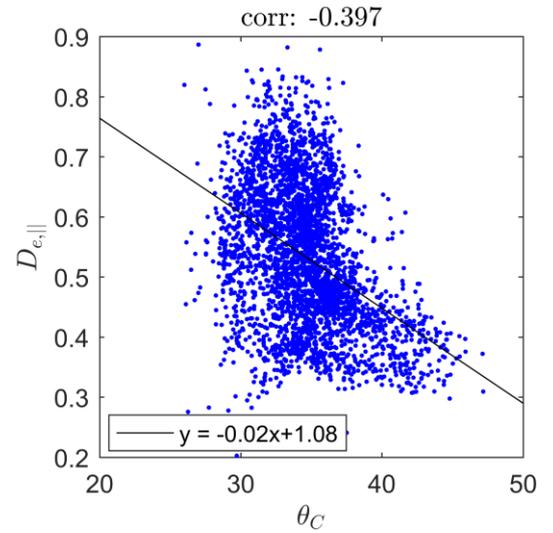

Fig. 7: Correlation between $\theta_C$ (in degrees) and WMTI estimates of $D_a$ and $D_{e,\parallel}$ for both branches in rat SC WM. The black line shows a robustfit to the data.

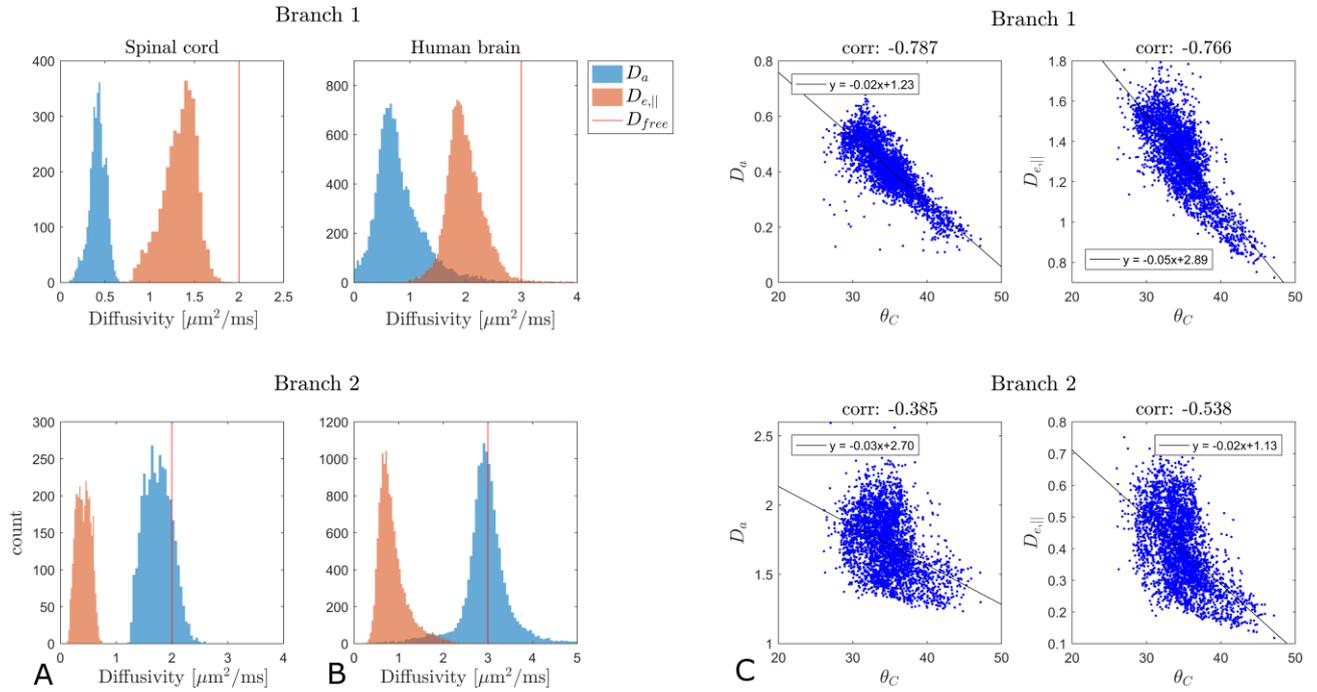

Fig. 8: aWMTI branch behavior in rat SC (A), human brain (B). Vertical red lines mark the free water diffusivity at the sample temperature. Panel C shows correlations between $\theta_C$ (in degrees) and aWMTI estimates of $D_a$ and $D_{e,\|}$ for both branches in rat SC WM. The black line shows a robustfit to the data.

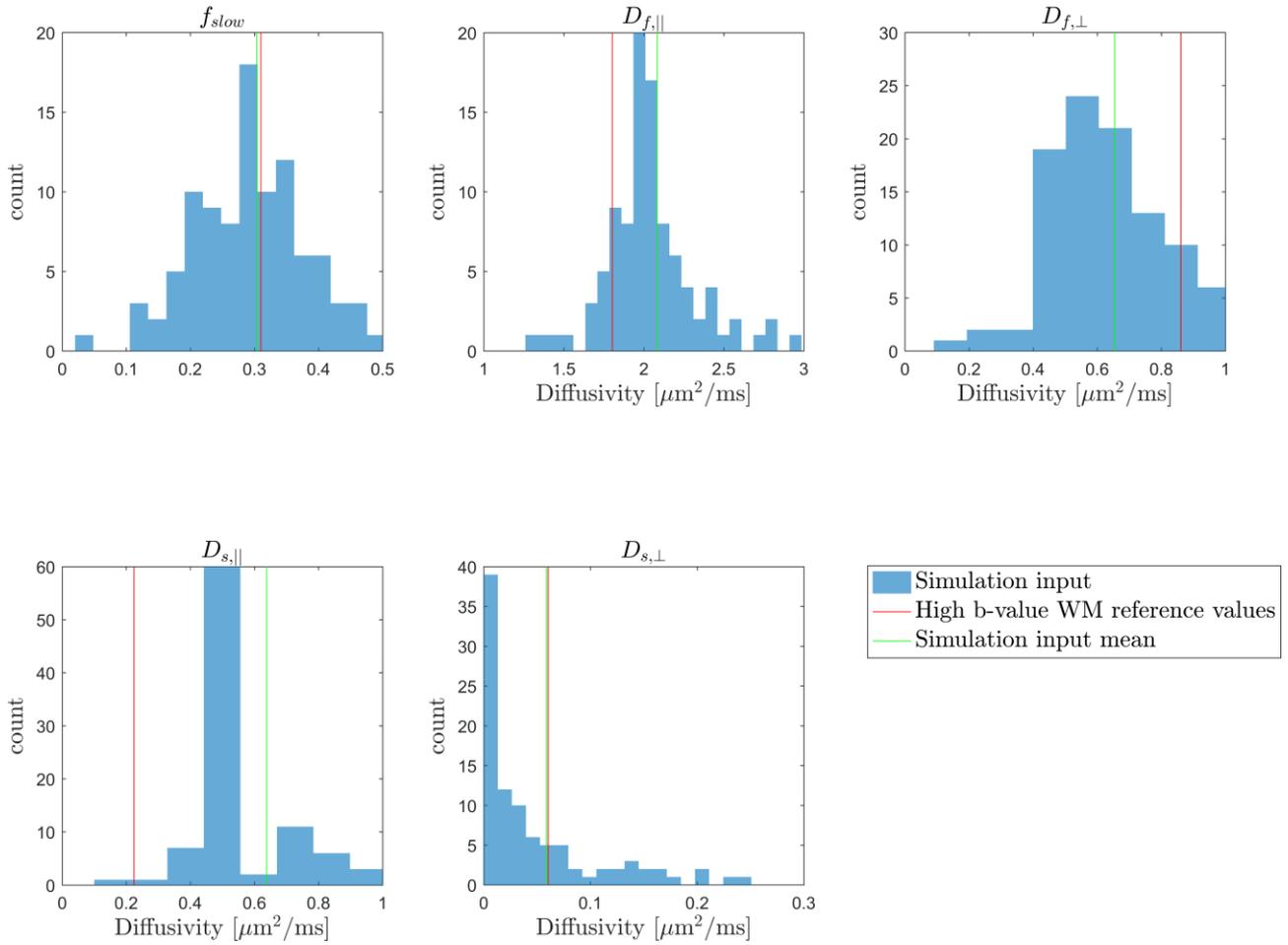

Supplementary Fig. 1: Histograms of biexponential parameter values from the 100 WM pixels used as input for the numerical simulations. We group the parameters as the slow (s) diffusion component, fast (f) diffusion component and the volume fraction of the slow component, $f_{slow}$. The green vertical line shows the distribution average for each parameter. Vertical red lines indicate corresponding parameter values obtained from human WM from high quality reference data to high b-value (47).

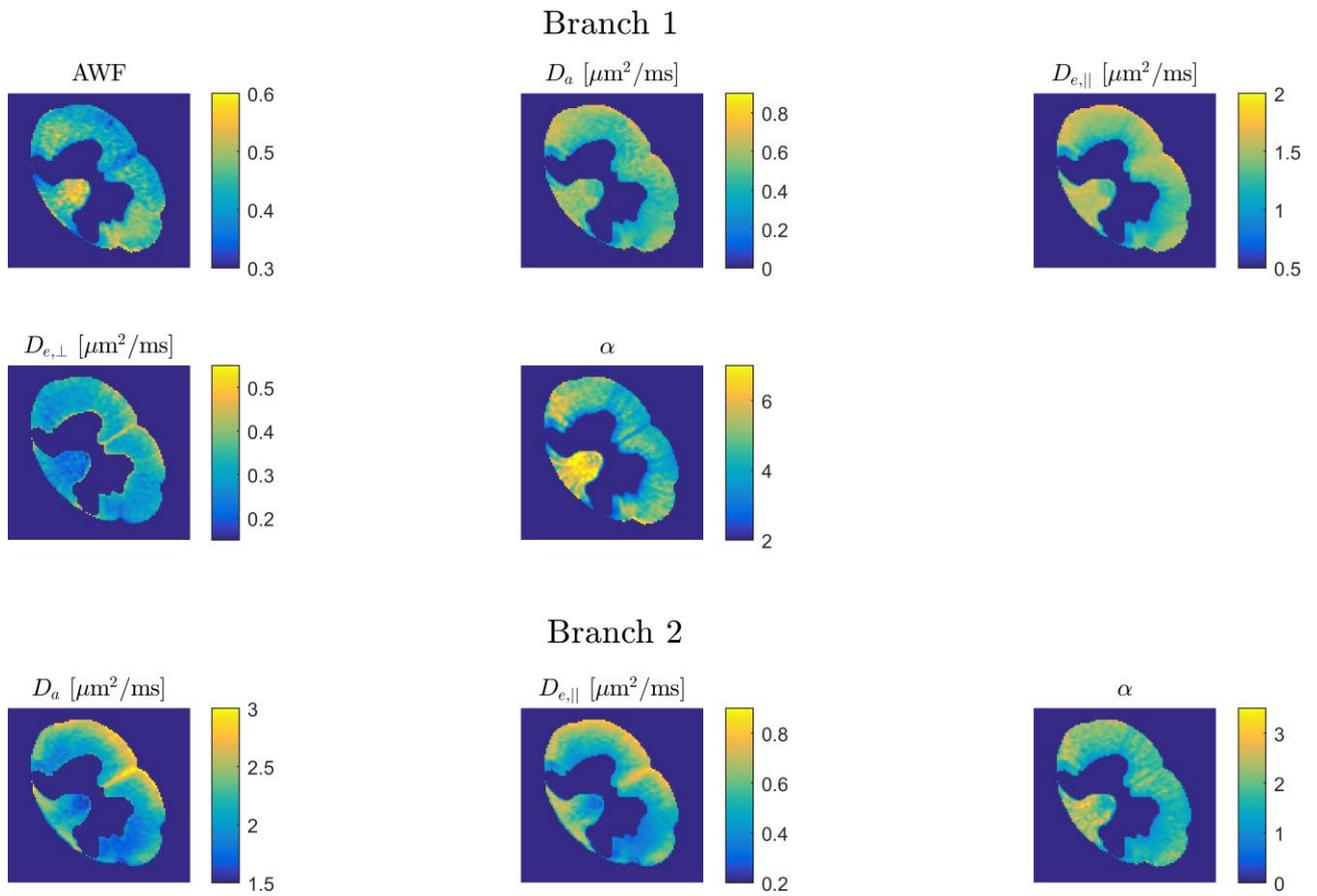

Supplementary Fig. 2: Maps of the five WM parameters in rat SC as estimated with WMTI.

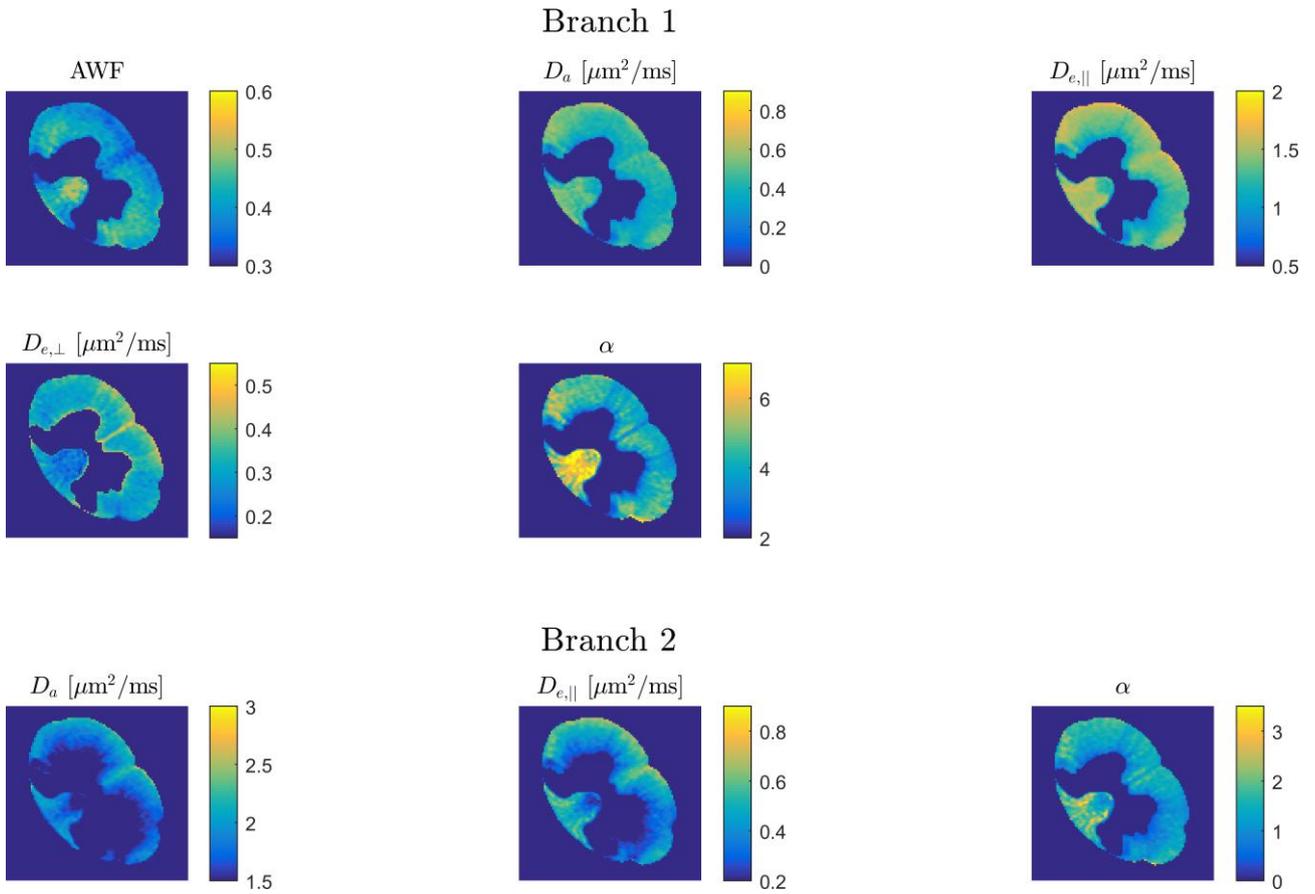

Supplementary Fig. 3: Maps of the five WM parameters in rat SC as estimated with aWMTI.

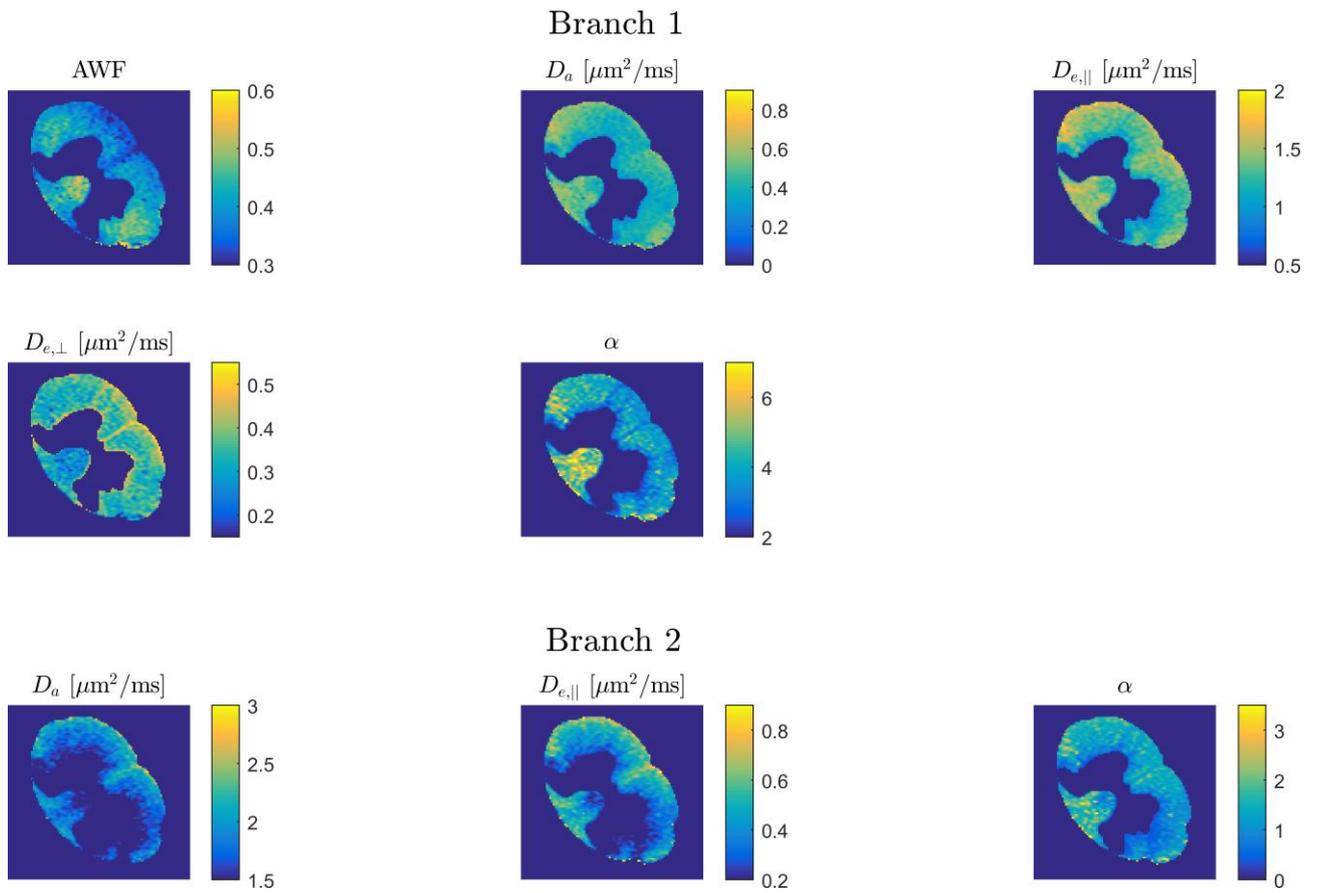

Supplementary Fig. 4: Maps of the five WM parameters in rat SC as estimated with faWMTI.